\documentclass{article}
\usepackage{amssymb}
\usepackage{authblk}
\usepackage{hyperref}

\title{Exciting Implications of LHC Higgs Boson Data\thanks{
		Sakurai Prize talk at the American Physical Society Meeting, Washington
		D.C., January 2017.}}
\author{Gordon Kane}
\affil{\it{Leinweber Center for Theoretical Physics (LCTP),}\\ \it{Department of Physics, University of Michigan}, \\ \it{Ann Arbor, MI 48109 USA}}

\begin{document}

\maketitle

\begin{abstract}
Naively, the LHC Higgs boson looks like a Standard Model Higgs boson, with
no guidance to physics beyond the Standard Model, as has often been
remarked. \ The data show that what was discovered is the true Higgs boson.
\ If one includes the full information available, experimental and
theoretical, there are actually four significant clues implied by data. \
They point toward a supersymmetric two-doublet decoupling theory, and a
hierarchy problem solution via TeV scale supersymmetry. That in turn
suggests an underlying compactified string/M theory with a de Sitter vacuum,
so we can be confident that the low scale model has an ultraviolet
completion.
\end{abstract}

\section{INTRODUCTION}
\indent

Nature has played an amusing and challenging trick on us. \ The Higgs boson
discovered at LHC in 2012 seems to be a Standard Model one, if one looks at
it's decay branching ratios (at the current level of precision). \ But of
course it cannot be a Standard Model one because of the hierarchy problem.
Fortunately, \ there is a well-known supersymmetric model that looks like
the Standard Model.\ For those who actually want to figure out what the
Higgs boson is telling us, there are four major clues.

Before listing the clues, it is worth making some remarks. Most importantly,
we know that what was discovered is the actual Higgs boson, because the
decays h$\rightarrow ZZ$ and $h\rightarrow W^{+}W^{-}$ are observed at full
strength. \ Both decays are forbidden in the Standard Mode, (because the $Z$
and $W$ are in electroweak triplet states, while $h$ is in an electroweak
doublet, and two triplet states cannot be combined to make a doublet). \ The
true vertex must be $hhZZ$ or $hhWW,$ with one $h$ getting a vacuum
expectation value (vev). \ The production rates are also full strength. \
Thus we learn that the Higgs mechanism \textit{is} operating, and the size
of the Higgs vev is full strength. \ None of the vev is distributed among
other Higgs states. \ Of course, the data has error bars, so the
quantitative remarks should be qualified by waiting for the data on $ZZ$ and 
$WW$ branching ratios to improve.

It did not have to come out that way. There could have been several Higgs
states sharing the Higgs vacuum values. Perhaps they would have been
directly seen, or the $ZZ$ or $WW$ branching ratios would have been somewhat
smaller. Current error bars still allow some sharing.\ There could have been
light HIggs partners observed. \ 

Clue 1. \ In the minimal supersymmetric world there is an upper limit on the
Higgs boson mass of at most about 130 GeV, which is satisfied for the
observed Higgs mass. \ The tree level lightest eigenstate is less than $%
M_{Z} $ and with top loop radiative corrections its mass increases up to
about 130 GeV. The observed HIggs boson mass is indeed lighter than that
limit.

Clue 2. In a supersymmetry world with low scale superpartners the hierarchy
problem is solved. That would hold here if gauginos were around the TeV
scale. \ That is still a possible result.

Clue 3. The well-known model [1,2] with large soft Higgs mass terms and two
Higgs doublets, satisfying the electroweak symmetry breaking conditions, has
one light Higgs eigenstate, two heavy neutral states, and a heavy charged
pair. \ It automatically has decay branching ratios that are very close to
the Standard Model ones, just as the data does. This is called the
decoupling solution, and has been familiar for over two decades. \ Such a
solution arises naturally in some UV complete theories, as we will briefly
discuss below.

Clue 4. The fourth clue is more subtle. \ For a single Standard Model Higgs
boson the effective Higgs potential is $V=\mu ^{2}h^{2}+\lambda h^{4}.$ In
the Standard Model $\lambda $ can run to go negative at larger scales, so
the potential becomes unbounded from below, and there is no resulting world.
\ Most people \ reacting to this situation have shrugged and said probably
the universe would be long lived so the instability can be ignored. \ But it
was pointed out [3-7] that without vacuum stability, fluctuations in the
Higgs field during inflation and in the hot early universe would have taken
most of the universe into an anti-De Sitter phase, giving a massive
collapse, and the expansion of the universe would never have occurred. The
point was basically raised explicitly in 2008, and there was some
uncertainty in how to properly calculate, over several years. \ Probably it
was settled by the significant paper [7] in 2017. \ The result is that for
generic expectations for the Hubble parameter during inflation, the Higgs
field fluctuations generated during inflation, or the \ hot, high density
early universe, probe the instability region, take \ most of the universe
into the unstable AdS phase, so the usual expansion of the universe fails to
occur. \ Thus the message is that the apparent instability is not
acceptable, and new physics must arise to stabilize the vacuum. \ In
supersymmetry $\lambda $ is determined by the gauge couplings ($\lambda
=(g_{1}^{2}+g_{2}^{2})/8),$ and is positive definite, so the vacuum
stability is automatic.

\section{IMPLICATIONS - LOW SCALE\ THEORY}

\indent

First, it is important to realize that there is a phenomenological Higgs
sector that behaves in a way consistent with the above clues. \ It is the
Higgs sector expected in a two Higgs doublet supersymmetric world, and has
long been known [1,2] to have these features. In a world with the terms of
the soft-breaking Lagrangian large, and thus the terms of the Higgs
potential large, and also the electroweak symmetry breaking conditions
(needed for allowing mass and for describing Z and W masses correctly)
satisfied, the Higgs boson mass is calculable and the Higgs decay branching
ratios are predicted to be very close to the Standard Model ones. Actually,
it is the ratio of the Higgs boson mass to the Z mass that is calculable
with precision. The electroweak symmetry breaking conditions for two
doublets implies that $M_{h}\lesssim M_{Z}.$The one top loop quantum
corrections raise the Higgs boson mass to about 125 GeV for heavy
soft-breaking terms. \ In the supersymmetric case the coefficient of the
quadratic term in the potential, $\lambda ,$ is calculable in terms of the
gauge couplings and is positive definite, so there is no vacuum stability
issue. \ The \ hierarchy problem is solved as usual in a supersymmetric
theory. It's \ well known that the two doublets rearrange into one effective
doublet that decouples, and one that is like the Standard Model doublet. \ 

It's very encouraging that a simple low-scale model exists that describes
the data well without any adjustable parameters, reported before the LHC
data [8]. It predicts four extra states, a heavy charged pair, a second
heavier CP even state, and a heavier CP odd state. \ None of them could be
detectable at planned electron-positron colliders or LHC upgrades, but they
could probably be seen at a future hadron collider with an energy several
times that of LHC. The model predicts that there are deviations from the
Standard Model h decay branching ratios, but they are at most a few per cent
in size, coming from chargino loops, so they are probably too small to
detect. \ 

The value of the Higgs boson mass is an interesting topic. \ The mass is
measured very well ($M_{h}\approx 125.11\pm 0.35GeV$) depending on how one
combines errors. That is better than one can ever hope to calculate it
theoretically. \ When we calculate it [8] we first work at the high scale,
string scale or unification scale. We calculate the soft breaking Lagrangian
and the resulting HIggs potential, \ and therefore the coefficient $\lambda $
of \ the quartic term. \ We match that to the effective theory at an
appropriate scale, e.g. the geometric mean of the stop mass eigenstates,
integrate out heavier scalar states, and run to the top scale. The squark
masses are required to be a few tens of TeV by the compactified theory.\
People \ have calculated two loop threshold corrections, and three loop beta
functions. \ We \ had a small workshop in December 3013 at the MCTP with
some experts to study how to do the matching and running, particularly with
the heavy scalars expected in good theories. Perhaps one can hope to
calculate the Higgs mass from a theory to nearly $1\%$ eventually, but that
is optimistic. There will be an additional scale uncertainty of about a
percent from doubling or halving the gravitino mass. \ Because \ the Higgs
mass \ is written at the compactification scale, followed by the running to
the low scale, it is not possible to show a simple, elegant formula for M$%
_{h}.$ 

Given \ the decoupling model and vacuum stability, one could ask if there
was significance to finding ourselves in the metastable region. \ The value
of $M_{h}$ is fixed by the electroweak breaking and the large soft terms
implied by the compactified theory. \ The top \ mass is fixed by the Higgs
vev and \ the top Yukawa coupling. \ The Yukawa coupling is determined by
the superpotential, a completely separate part of the theory. \ There is not
yet an understanding of why there should be one and only one large quark
Yukawa coupling, but it seems that beieng in the \ metastable region is a
coincidence.

\section{ULTRAVIOLET COMPLETION?}
\indent

It is thought that almost all low scale models for Higgs sectors, dark
matter, LHC physics, etc., do not have a ultraviolet (UV) completion. \ That
is, they could not have originated from a theory that included a quantum
theory of gravity. \ Such models/theories are said by Cumran Vafa to live in
the "swampland". \ Vafa in his TASI lectures [9] suggests some criteria, but
it is hard to know if a low scale model is in the swampland. \ By far the
best way to be confident that a low scale model has a UV completion, which
is of course necessary if we hope it is relevant to describing our world, is
to start with a 10 or 11 dimensional string/M-theory, and compactify it to 4
dimensions, .

So we ask if our two-doublet HIggs sector with large soft-breaking terms
arises from a compactified string/M theory? \ The answer is yes, it has been
demonstrated [10] that compactifying M-theory on a manifold with G$_{2}$
holonomy necessarily gives a 4-dimensional supersymmetric quantum field
theory. Moduli are all stabilized, in a de Sitter vacuum. Vafa [9] states
that the resulting vacuum is not de Sitter, which is true for his
assumptions. \ But including hidden sector charged matter (which is
generically present and which he did not include) gives a deS vacuum, via
the additional F-terms - their contribution cancels the -3W\symbol{94}2 term
in the scalar potential, giving a positive energy [10]. So the full theory
does have a de Sitter vacuum.

The resulting theory automatically implies the supersymmetry is
softly-broken via gravity mediation, and generically satisfies the
conditions for electroweak symmetry breaking, with a Higgs mass about equal
to the observed one. \ The compactified M-theory has been shown to have
Yang-Mills gauge theories like the Standard Model [11] and chiral quarks and
leptons like the Standard Model [12]. \ It solves the hierarcy problem, and
it can support grand unification [13] and has axions, and a solution to the
strong CP problem [14].

It has two dark matter candidates, axions and hidden sector stable matter.\
The theory does not have adjustable parameters, though our present inability
to calculate some things means some results are poorly known. With all these
successful tests we can be rather confident that our Higgs sector qualifies
as a realistic one.\ One such example is sufficient to be satisfied that the
low scale description can arise in an underlying theory that is consistent
with also \ having quantum gravity [15]. \ 

\subsection{FINAL REMARKS - A DREAM FULLFILED, NOT A NIGHTMARE}
\indent

Old fashioned people could view the Higgs physics in a traditional way - the
compactified M-theory predicted the Higgs boson and its mass (ratio to the Z
mass) and its decay branching ratios, without free parameters, and
experiment has confirmed them, which helps validate the theory. The
compactified theory also predicts additional tests: it predicts that gluinos
are in the LHC range before any major upgrades if it collects enough
luminosity, that the g$_{\mu }-2$ experiment at Fermilab should not see
significant deviations from the Standard Model, that electric dipole moments
are smaller than the current limits but not much smaller [16], and more.

People sometimes say that the situation after LHC could be the nightmare one
with only a Standard Model Higgs boson and no guidance as to how to proceed.
Sometimes people say Higgs physics is \textit{a mystery}. Those statements
are \textit{wrong}, and arise from ignorance about or ignoring information
and clues. \ It is \textit{not surprising} that the Higgs boson branching
ratios are like the Standard Model ones, because such a situation occurs
naturally in well-known robust models, and those models have possible
ultraviolet completions (as a \ model must to not be in the swampland). The
Higgs boson whose field breaks the electroweak symmetry has been found, and
the data do provide information that is helpful in guiding us toward how to
extend the Standard Model. \ Higgs physics is not a mystery.

\section*{Acknowledgments}

I'm grateful to Bibyushan Shakya and Malcolm Perry for helpful discussions,
and to Eric Gonzalez for help with the manuscript. \ This research was
supported in part by the Department of Energy. \ 

\section*{References}
\indent

1. \textit{Multiscalar Models With a High-energy Scale}, Howard E. Haber,
Yosef Nir, Nucl.Phys. B335 (1990) 363-394

2. \textit{The CP conserving two Higgs doublet model: The Approach to the
decoupling limit}, John F. Gunion, Howard E. Haber, Phys.Rev. D67 (2003)
075019, hep-ph/0207010

3.\textit{\ Cosmological implications of the Higgs mass measurement}, J.R.
Espinosa, G.F. Giudice, A. Riotto, JCAP 0805 (2008) 002, arXiv:0710.2484

4. \textit{The Probable Fate of the Standard Model}, J. Ellis, J.R.
Espinosa, G.F. Giudice, A. Hoecker, A. Riotto, Phys.Lett. B679 (2009)
369-375, arXiv:0906.0954

5. \textit{Probable or Improbable Universe? Correlating Electroweak Vacuum
Instability with the Scale of Inflation} A.Hook, J. Kearney, B. Shakya, and
K.M.Zurek, JHEP 1501 (2015)061, arXiv:1

6. \textit{The cosmological Higgstory of the vacuum instability}, Jose R.
Espinosa, Gian F. Giudice, Enrico Morgante, Antonio Riotto, Leonardo
Senatore, Alessandro Strumia , Nikolaos Tetradis, JHEP 1509 (2015) 174;
arXiv:1505.04825

7. \textit{Spacetime Dynamics of a Higgs Vacuum Instability During Inflation}%
, William E. East, John Kearney, Bibhushan Shakya, Hojin Yoo, Kathryn M.
Zurek, Phys.Rev. D95 (2017) no.2, 023526, arXiv:1607.00381

8. \textit{Higgs Mass Prediction for Realistic String/M Theory Vacua},
Gordon Kane, Piyush Kumar, Ran Lu, Bob Zheng, Phys.Rev. D85 (2012) 075026,
arXiv:1112.1059

9. \textit{The String Landscape, the Swampland, and the Missing Corner}, T.
Daniel Brennan, Federico Carta, Cumrun Vafa, arXiv:1711.00864

10. \textit{Explaining the Electroweak Scale and Stabilizing Moduli in M
Theory}, Bobby Samir Acharya, Konstantin Bobkov, Gordon L. Kane, Piyush
Kumar, Jing Shao, Phys.Rev. D76 (2007) 126010, hep-th/0701034

11. \textit{M theory, Joyce orbifolds and superYang-Mills}, Bobby Samir
Acharya, Adv.Theor.Math.Phys. 3 (1999) 227-248, hep-th/9812205

12. \textit{Chiral fermions from manifolds of G(2) holonomy}, Bobby Samir
Acharya, Edward Witten, hep-th/0109152

13.\textit{\ Deconstruction, G(2) holonomy, and doublet triplet splitting},
Edward Witten, hep-ph/0201018

14. \textit{An M Theory Solution to the Strong CP Problem and Constraints on
the Axiverse}, Bobby Samir Acharya, Konstantin Bobkov, Piyush Kumar, JHEP
1011 (2010) 105, arXiv:1004.5138

15. \textit{String Theory and the Real World}, Gordon Kane, Morgan
Claypool/British Instutute of Physics Short Book, 2017

16. \textit{Theoretical Prediction and Impact of Fundamental Electric Dipole
Moments}, Sebastian Ellis and Gordon Kane, JHEP 1601 (2016) 077,
arXiv:1405.7719

\end{document}